# The Role of Tag Suggestions in Folksonomies

Dirk Bollen

Faculty of Innovation Sciences and Industrial Engineering
University of Technology Eindhoven
Eindhoven, The Netherlands
d.g.f.m.bollen@tue.nl

Harry Halpin

Institute for Communicating and Collaborative Systems
University of Edinburgh
Edinburgh, United Kingdom
H.Halpin@ed.ac.uk

October 27, 2018

## Abstract

Most tagging systems support the user in the tag selection process by providing tag suggestions, or recommendations, based on a popularity measurement of tags other users provided when tagging the same resource. In this paper we investigate the influence of tag suggestions on the emergence of power law distributions as a result of collaborative tag behavior. Although previous research has already shown that power laws emerge in tagging systems, the cause of why power law distributions emerge is not understood empirically. The majority of theories and mathematical models of tagging found in the literature assume that the emergence of power laws in tagging systems is mainly driven by the imitation behavior of users when observing tag suggestions provided by the user interface of the tagging system. This imitation behavior leads to a feedback loop in which some tags are reinforced and get more popular which is also known as the 'rich get richer' or a preferential attachment model. We present experimental results that show that the power law distribution forms regardless of whether or not tag suggestions are presented to the users. Furthermore, we show that the real effect of tag suggestions is rather subtle; the resulting power law distribution is 'compressed' if tag suggestions are given to the user, resulting in a shorter long tail and a 'compressed' top of the power law distribution. The consequences of this experiment show that tag suggestions by themselves do not account for the formation of power law distributions in tagging systems.

# 1 Introduction

During the last decade the Web has become a space where increasing numbers of users create, share and store content, leading it to be viewed not only as an "information space" [2] but also a "social space" [9]. This new step in the evolution of the Web, often referred to as the "Web 2.0," was shaped by the arrival of the different services that came into existence to support users to easily publish content on the Web, such as photos (Flickr), bookmarks (del.icio.us), movies (YouTube), weblogging (Wordpress), and so on [12]. Almost simultaneously with the growth of user-generated content on the Web came a need create order in this fast growing unstructured data. Tagging has become the predominant method for organizing, searching and browsing online web resources[1] in this social web. Tagging refers to the labeling of web-resources by means of free-form descriptive keywords. With tagging users themselves annotate web-resources by tags they freely chose and thus forms a 'flat space of names' without the predefined and hierarchical structure characteristic of classic 'ontologies' in knowledge engineering.

Instead of traditional expert-defined taxonomies, the tagging of web resources presents an alternative decentralized and user-generated categorization referred to as a *folksonomy*. The term 'folksonomy' is itself a combination of 'folk' and 'taxonomy' [19]. The advocates of folksonomies usually claim that, as opposed to taxonomies, folksonomies are a "social" classification process in which tags and content are shared with other users [10]. The most general claim is that the collective tagging of web resources gives rise to the emergence of a user-driven vocabulary that users can share and that is more flexible than traditional classification methods, as Shirky states that knowledge engineers tend to "overestimate the amount to which users will agree" [16]. However, work by Halpin et al. [8] clearly undermines Shirky's presupposition that users will not agree on tags by showing that based on actual tagging data, users tend to agree on many tags for a given resource. What precisely is the process that leads to users repeating tags and agree on tags? Can these factors be understood and modelled correctly? If not, how can models be improved? These questions are not trivial and experiments in the style of cognitive psychology can help.

## 1.1 Folksonomies: Stable or Not?

Tagging often gets criticized because it is too chaotic since the ordinary user in tagging systems is not trained for performing categorization, and is therefore assumed likely to be beset by a host of problems that a professional knowledge engineer might avoid in categorizing knowledge. Many of these problems are syntactic, such as misspellings, or are based on cultural differences, such as the proliferation of tags in different languages. However, these syntactic problems can be be solved by the use of (multi-lingual) dictionaries and tag completion, although these features are not widely deployed in existing tag-systems. More serious difficulties of tagging systems originate from their inability to provide the user anything beyond word-based semantics for tags, exemplified by the use of semantically ambiguous tags and the inability to express structured data like dates in tags, as in "19 July 2008." Recent advances in tag-systems show that relatively simple methods from natural language processing and assumptions (such as that the user generally uses one 'sense' of an ambiguous tag) can resolve these issues [1].

---

[1] A web resource is anything that can be given a URI (Uniform Resource Identifier, including but not limited to web-pages [2].

The most serious allegation leveled either in favor or against tagging systems is that due to lack of a centralized vocabulary the users of the tagging system will never manage to converge their tags to a stable collective categorization scheme needed to describe a resource [16]. Users may differ in viewpoints, purposes, and sociocultural backgrounds about that resource, so the same resource could be tagged by different persons with distinct tags [10]. However, it could be hypothesized that an emergent collective description of a resource will arise from the decentralized tagging behavior, since a resource can receive hundreds to thousands tags, eventually certain tags will stand out because they received most tag entries and present a 'consensus' on how to describe a certain resource.

Empirical studies of del.icio.us show that the number of tags needed to describe a resource consistently converges to a power law distribution as a function of how many tags it receives [7]. We refer to the highest ranked frequencies of the power-law distribution as the 'top' of the distribution, as opposed to the long tail. Furthermore, we can consider the formation of a power law distribution to be 'stable' due to its well-known property of true power law generating functions known as *scale invariance*. A power law distribution produced by tagging is a good sign of stability since, due to scale invariance, increasing the number of tagging instances only proportionally increases the scale of the power-law, but does not change the parameters of the power law distribution. Thus, the first step in determining if users have reached a stable consensus in tagging is the detection of a power law distribution from the frequencies of tags.

Regarding the tag frequency distributions as probability distributions, this process of stabilization can be detected by the use of Kullback-Leibler Divergence, a information-theoretic metric that describes the differences between two probability distributions [8]. There are some odd features to the power law distribution produced by users tagging a single resource, in particular the presence of a 'bump' that lengthens the top of the distribution until the seventh to tenth tag. This 'bump' has been thought to possibly be an artifact of the user interface that provides tag suggestions to users, since the user interface of del.icio.us usually provides tag suggestion on up to a maximum of ten tags [8]. The empirical results by Halpin et al. show that far from being unstable and chaotic, tagging systems are in fact incredibly stable and stabilize relatively quickly [8], with users able to reach consensus on a small amount of heavily repeated 'core' terms in a vocabulary, given by the top of the distribution, with a long-tail of more idiosyncratic terms. Once these tags have stabilized, these tags can then reliably consumed by other applications, such as their use with algorithms from network analysis to detect communities [13, 15].

The reasons behind tag stabilization and the emergence of a power law are yet unknown, although explanations fall into two general categories. The first of these explanations is relatively simple: the tags stabilize because users are imitating each other via tag suggestions put forward by the tagging system [7]. The second and more recent explanation is that in addition to imitation, the users share the same background knowledge [6]. In order to draw apart the relative influence of the imitation based on feedback from shared background knowledge. It is important to elicit the processes that lead to the emergence of the power law, as a mere power law distribution can be generated by a number of different models, many incredibly simple and unrealistic. Furthermore, it is also important to accurately determine if the data is indeed a power-law, as most inspections of tagging data for power-laws has been done visually, which is well-known to be unreliable. What is necessary for a scientific explanation of tagging systems is that the model being proposed actually provide an account of the actual informational and cognitive behavior of tagging users. A model that only generates a power law distribution is not enough.

## 1.2 Related work

The role of tag suggestions on tagging has been studied extensively. Sen et al. use a survey-based approach in general, but also states that tag suggestions "indirectly" influence tagging behavior, although this is shown via a cosine similarity metric of current tags to previous tags seen [14]. Despite their hypothesis, this similarity function does not seem to vary much regardless of tags seen and is compared to an unrealistic uniform tag distribution as a baseline, that does not take into account the possibility of shared background knowledge [14]. While they do study not displaying tags at all to users when comparing different tag suggestion display algorithms, and they suggest that 'personal' tags for users without tag suggestions do not converge [14]. However, Sen et al. do not take into account that the mass of a power law distribution is in the long tail regardless, so users without tag suggestions could converge to power law distributions that share the top of the distribution while still employing 'personal' tags in the long tail.

[18] compared different tag suggestion algorithms in a Web-based experiment. They developed two metrics, i.e. matching rate and imitation rate

to determine the influence of suggestion on user tag applications. Results of this study show that users are influenced by tag suggestions, since they show that an average 1 out of 3 tags were selected from the suggested tags when provided by the tagging system. However, the vast majority of users in their system noted that "In general I pay no attention to suggested tags" and only 10% found them "helpful" [18]. Although the experiments performed by [14] and [18] study the role of tag suggestion on tag behavior, neither experiment performed an analysis of the effects of tag suggestions on the emergence of power law distributions of tag frequencies over resources. While these studies look at the influence of tag suggestion on individual tag behavior (e.g. the reuse of tags due to tag suggestions, they do not investigate or assume a model underlying tag behavior.

### 1.3 Outline

The outline of this paper is as follows. In Section 2 we re-iterate the widely used definitions and formal models of tagging. Section 3 describes the user study performed to determine the relative influence of background knowledge and tag suggestions on the emergence of a power distribution for tagged resources. Finally, Section 4 presents the results of the user study.

## 2 Models of collaborative tag behavior

### 2.1 Formalizing Tagging

The traditional tripartite model of tagging is well-known. In essence, in a *tagging instance* a user $u$ applies $n$ tags $(t_1...t_n)$ in order to categorize a given resource $r$. So, a tagging instance $p$ can be identified as the triple $p = (u, r, (t_1...t_n))$. Since these tagging instances are given over time, one can identify a *tagging stream* $m$ as a time-ordered series of tagging instances over time (dates) $d_1...d_j$. There are three metrics that are often used to describe tagging systems. The first is the *tag-resource distribution*, which inspects the frequency that each tag $t_1....t_k$ has been applied to a given resource $r$ by a number of distinct users $u_1...u_x$. Therefore, in tag-resource distributions, each tag is assigned a frequency $f$, the number of times the tag has been repeated for a particular resource. In general, when we are referring to a distribution we are referring to the tag-resource distribution. This distribution is graphed by ordering the tags $t_1...t_k$ in rank order on the $x$ axis against their frequency on the $y$ axis, with the highest frequency first and the rest in descending order. Further metrics that are of interest to researchers are *tag-growth distributions*, which counts the number of distinct tags $t$ assignments over some period of time over all users and resources in a tagging system. Another distribution is the *tag-correlation distributions*, which is the tag frequency for two tags $t_i$ and $t_j$ occurring in the same tagging instance $p$ for all tags in the tagging system.

## 2.2 A simple model: The Polya Urn

The most elementary model of how a user selects tags when annotating a resource is simple imitation of other users. Note that 'imitation' in tagging systems means that the tags are being reinforced via a 'tag suggestion' mechanism, and so the terms "imitation", "reinforcement", "feedback", and 'tag suggestion' can be considered to be synonymous in the context of tagging systems. The user can imitate other users precisely because the tagging systems tries to support the user in the tag selection process by providing tag recommendations based on tags other people used when tagging the same resource. There are minor variants of this theme, such as the possibility of using a combination of tags of other users in combination with a user's own previously used tags. In most tagging systems like del.icio.us these tag suggestions are presented as a list of tags that the user can select in order to add them to their tagging instance. The selections of tags from the tag recommendation forms a positive feedback loop in which more frequent tags are being reinforced, thus causing an increase in their popularity, which in turn causes them to be reinforced further and exposed to ever greater numbers of users. This simple type of explanation is easily amendable to preferential attachment models, also known as 'rich get richer' explanations, which are well-known to produce power law distributions. Intuitively, the earliest studies of tagging observed that users imitate other pre-existing tags [7]. Golder and Huberman proposed that the simplest model that results in a "power law" would be the classical Polya urn model [7]. Imagine that there is urn containing balls, each of some finite number of colors. At every time-step, a ball is chosen at random. Once a ball is chosen, it is put back in the urn along with another ball of the same color, which formalizes the process of feedback given by tag suggestions. As put by Golder and Huberman, "replacement of a ball with another ball of the same color can be seen as a kind of imitation" where each color of a ball is made equal to a natural language ta and since "the interface through which users add bookmarks shows users the tags most commonly used by others who bookmarked that URL already; users can easily select those tags for use in their own bookmarks, thus imitating the choices of previous users" [7]. Yet, this model is too limited to describe tagging, as it features only reinforcement of existing tags, not the addition of *new* tags.

## 2.3 Imitation and The Yule-Simon Model

The first model that formalized the notion of new tags was proposed by Cattuto et al. [4]. In order for new tags to be added, a single parameter $p$ must be added to the model, which represents the probability of a new tag being added, with the probability $\bar{p} = (1 - p)$ that an already-existing tag is reinforced by random uniform choice over all already-existing tags. This results in a Yule-Simon model, a model first employed by Yule [20] to model biological genera and later Simon to model the construction of a text as a stream of words [17]. Furthermore, while not assuming any frequency distribution at all for tags, this model results in a power law for the rate of tag-growth distributions, whose exponent is $\alpha \sim 1 - p$ ($P(k) \sim k^{-\alpha}$ with $\alpha = 1 + 1/\bar{p}$). This model has been shown to be equivalent to the famous Barabasi and Albert algorithm for growing networks [3]. Yet the standard Yule-Simon process does not model vocabulary growth in tagging systems very well, as noticed by Cattuto et al. as it produces exponents "lower than the exponents we observe in actual data" [4].

Cattuto et al. hypothesize that this is because the Yule-Simon model assumes users are choosing to reinforce ($\bar{p}$) tags uniformly from a distribution of *all* tags that have been used previously, so Cattuto concludes that "it seems more realistic to assume that users tend to apply recently added tags more frequently than old ones" [4]. This behavior could be caused by the exposure of a user to a feedback mechanism, such as del.icio.us tag suggestion system. This suggestions exposes the user only to a subset of previously existing tags, such as those most recently added. Since the tag suggestion mechanism only encourages more recently-added tags to be re-enforced with a higher probability, Cattuto et al. added a memory kernel with a power law exponent to standard Yule-Simon model. This means that the weight of a previously existing tag being reinforced is weighted according to a power law itself, so that a tag that has been applied $x$ steps in the past is chosen with a probability $Q_t(x) = a(t)/(x+\tau)$, where $a(t)$ is a normalization factor and $\tau$ "is a characteristic time scale over which recently added words have comparable probabilities." While the parameter $p$ controls the probability of reinforcing an existing tag, this second parameter $\tau$, controls how fast the memory kernel decays and so over what time-scale a tag may likely count as 'new' and so be more likely to be reinforced. As Cattuto et al. notes, "the average user is exposed to a few roughly equivalent top-ranked tags and this is translated mathematically into a low-rank cutoff of the power law" [4]. This model produces an "excellent agreement" with the results of tag-correlation graphs. It should be clear that the original Yule-Simon model, in its standard interpretation, just parametrizes the probability of imitation of existing tags, and the modified Yule-Simon model with a power law memory kernel also depends on imitation of existing tags, where the probability of a previously-used tag is just decaying according to a power law function.

## 2.4 Adding Parameters and Background Knowledge

Although Cattuto et al.'s model is without a doubt an elegant minimal model that captures tag-correlation distributions well, it was not tested against tag-resource distributions [4]. Furthermore, as noticed by Dellschaft and Staab, Cattuto et al.'s model also does not explain the sub-linear tag vocabulary growth of a tagging system [6]. Dellschaft and Staab propose an alternative model, which adds a number of new parameters that fit the data produced by tag-growth distributions and tag-resource distributions better than Cattuto et al.'s model [6]. The main points of interest in their model is that instead of new tag being chosen uniformly, the new tag is chosen from a power law distribution that is meant to approximate "background knowledge." However, their model also features as the inverse of "background knowledge" the "probability that a user imitates a previous tag assignment" [6]. In essence, Dellschaft and Staab have added (at least) two new parameters to a Yule-Simon process, and these additional parameters allows the reinforcement of existing tags to be more finely tuned. Instead of a single power law memory kernel with a single parameter $\tau$, these additional parameters allow the modeling of "an effect that is comparable to the fat-tailed access of the Yule-Simon model with memory" while keeping tag-growth sub-linear [6]. The model proposed by Cattuto et al. kept the tag-growth parameter equal to 1 and so makes tag growth linear to $p$ [4]. Yet for us, most important advantage of Dellschaft and Staab over Cattuto et al.'s model is that their added parameters lets their model match the previously unmatched observation by Halpin et al. of the frequency rank distribution of resources being a power law [8]. The match is not as close as the match with vocabulary growth and tag correlations, as resource-tag frequency distributions vary highly per resource, with the exception of the drop in slope around rank 7-10 [8].

## 2.5 Research Questions

What unifies all of these models is that they assume that tag suggestions from the tagging system has a major impact on the emergence of a power law distribution. With concern to the modified Yule-Simon model and the more highly parametrized model that takes into account 'background knowledge,' different claims are made of where the imitated tags come from. Cattuto et al. proposes that they come from a random uniform distribution of tags while Dellschaft and Staab propose a more topic-related distribution that itself has a power law distribution [6]. However, just because a simple model based on imitation of tag suggestions can lead to a power law distribution does not necessarily mean that tag suggestions are actually the causal mechanism that causes the power law distribution to arise in tagging systems.

The research questions posed then are: (a) Is imitation behavior, and therefore the tag suggestion mechanism, the main force behind the observed power law distributions in tagging systems? (b) What is the role of tag suggestions on the total tag-resource distribution?

## 3 Experimental Design

In order to measure the effects of tag suggestions on the tag behavior of users we developed a web based experiment in which participants were asked to tag 11 websites, with two varying conditions: the 'tag suggestion' condition (Condition A) in which 7 tag suggestions were presented to the participant, and a 'no tag suggestion' condition (Condition B) in which no tag suggestions were presented to user.

In this experiment we focus on del.icio.us which is the most known and widely used social tagging systems. Del.icio.us was the first to introduce a tag based collaborative bookmark system. Del.icio.us has more than five million users and 150 million tagged URIs and so provides a vast data-set which makes it the most studied Web 2.0 services. The user interface used in our experiment presented the tag suggestions in a similar way to del.icio.us to avoid confusion.

The 11 websites were selected according to two criteria. First, the topics of the websites needed to appeal to a general public. Second, the website needed to have over 200 tagging instances. The appeal to the general public was operationalized by randomly choosing sites that were tagged with the tag "lifestyle" on del.icio.us. The tag "lifestyle" is a popular tag with 72,889 tagged resources as of October 2008. This was chosen in order to not bias our study to one particular specialized subject matter, and so exclude resources on del.icio.us that have a highly technical content. Specialized content may not lead to normal tagging behavior from participants in the experiment, who might not be familiar with the subject matter. The second criteria of using only resources with over 200 tagging instances was chosen since it has been shown that stable power law tag distributions emerge around the 100-150th tagging [7]. We did not want the tag suggestions to be from non-stable tag distributions, as it has been shown that the variance between the top popular tag could vary widely before 100-150th tag. The 11 websites selected for this experiment, with the popular tags provided from del.icio.us and the number tags. Note that while the number of URIs 11 may appear to be small, it is larger than previous experiments over tag suggestions [18] and was enough to give the experiment enough power to be statistically significant. It was far more critical for this experiment to get enough subjects in order for power-law distributions to be given the chance to arise without tag suggestion, and this would require at least 100 experimental subjects tagging each URI.

Figure 1 shows the experimental design. In the 'no tag suggestion' condi-tion (Condition A), as shown in Figure 1, a user is presented the 11 websites he needs to tag without any form of tag suggestions. In the 'tag suggestion' condition (Condition B), also shown in Figure1, a user is presented the 11 websites with 7 suggested tags. While the details of the tag suggestion algorithm applied by del.icio.us is unknown, for our experiment the suggested tags in condition B were aggregated from delicious and are the top 7 popular tags for each of the 11 websites. These popular tags are the 7 most frequent used tags for a particular website provided by users over the tagging history of the resource. For the experiment the 7 popular tags were aggregated and presented to the participants in manner similar to how tags are suggested to users of del.icio.us, being shown to the user before they commence their tagging. Each of the 300 participants was randomly assigned to either the 'tag suggestion' or 'no tag suggestion' condition. From these 300 participants 78 did not tag any website (37 in the 'tag suggestion' condition, 41 in the 'tag suggestion' condition) and are therefore excluded from further analysis. The participants were randomized over age, gender, computer, Internet and tag use.

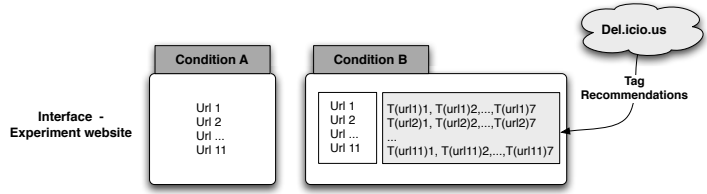

Figure 1: Experimental Design

## 4 Results

In total the 222 participants applied 7,250 tags over all websites in both conditions. with 3,694 tags applied in the 'tag suggestion' condition and 3556 in the 'no tag suggestion' condition. On average every user in the 'tag suggestion' condition applied 32.69 ($S.D. = 9.77$) tags over all 11 URIs and for the no tag suggestion conditions 32.61 ($S.D. = 6.80$) tags over 11 URIs.

### 4.1 Detecting Power Law Distributions

Since the primary goal of this experiment is to investigate the role of tag suggestion on the emergence of a power in collaborative tagging systems we plotted the ranked frequency distribution of all websites in both condi-tions in Figure 2. Figure 2 depicts all tag-resource distributions for all 11 experimental websites for both conditions.

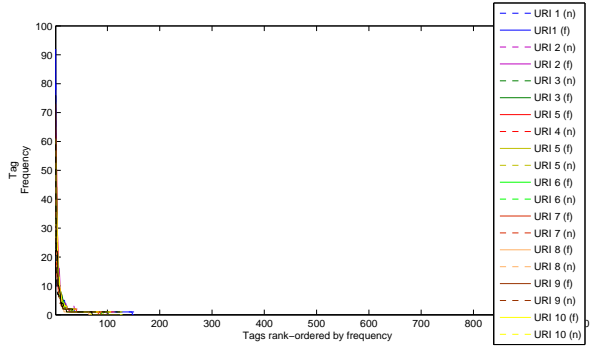

Figure 2: Depicts the tag-resource distributions for the 11 websites in the experiment. The different colors represent the tag-resource distribution of different URIs in the experiment. The 'tag suggestion' (f) condition is given as a solid line, while the 'no tag suggestion' (n) condition is given as a dotted line.

The power law distribution is defined by the function:

$$y = cx^{-\alpha} + b \qquad (1)$$

in which $C$ and $\alpha$ are the constants that characterize the power law and $b$ being some constant or variable dependent on $x$ that becomes constant asymptotically. The $\alpha$ exponent is the scaling exponent that determines the slope of the distribution before the long tail behavior begins. A power law function can be transformed to a log-log scale as in the following equation:

$$log(y) = -\alpha log(x) + log(C) \qquad (2)$$

This equations shows the characteristic properties of power law function is that when transformed to a log-log scale the power law distribution takes the shape of a linear function with slope $\alpha$. So transforming a function to a log-log scale and determining the slope $\alpha$ is one of the first steps in examining if a distribution has a power law. We plotted all 11 tag-resource distributions in log-log space in Figure 3. From this first look on the data it seems that power laws emerge in both the 'tag suggestion' and 'no tag suggestion' conditions. In order to clarify their differences, we averaged the tag-resource distributions, and this is given in Figure 4. In a log-log scale, *both* conditions appear visually to exhibit power law behavior.

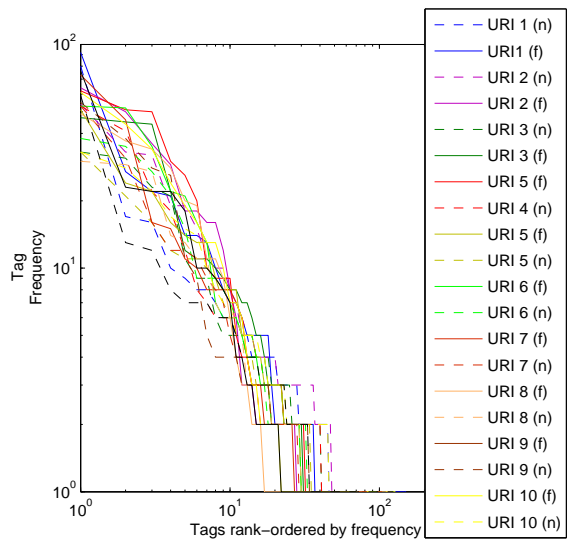

Figure 3: Depicts the tag-resource distributions for the 11 websites in the experiment on the log-log scale. The different colors represent the tag-resource distribution of different URIs in the experiment. The 'tag suggestion' (f) condition is given as a solid line, while the 'no tag suggestion' (n) condition is given as a dotted line.

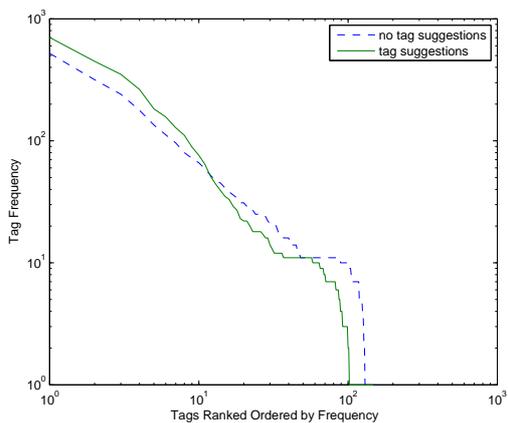

Figure 4: Averaged tag-resource distributions for both experimental conditions on a log-log scale. The solid line depicts the 'tag suggestion' condition, the dotted line the 'no tag suggestion' condition.
12

### 4.1.1 Parameter Estimation via Maximum-Likelihood

The most widely used method to check whether a distribution follows a power-law is to apply a logarithmic transformation, and then perform linear regression, estimating the slope of the function in logarithmic space to be $\alpha$. However, this least-square regression method has been shown to produce systematic bias, in particular due to fluctuations of the long tail [5]. To determine a power-law accurately requires minimizing the bias in the value of the scaling exponent and the beginning of the long tail via maximum likelihood estimation. See Newman [11] for the technical details. To determine the $\alpha$ of the observed distributions, we fitted the data using the maximum likelihood method recommended by Newman [11]. Figure 5 shows the different $\alpha$ parameters for the 'tag suggestion' and 'no tag suggestion' conditions, as well as the $\alpha$ aggregated from data from del.icio.us. Overall, for the 'no tag suggestion' condition, the average $\alpha$ was 2.1827 (S.D. 0.0799) while for the 'tag suggestion' condition the average $\alpha$ was 2.0682 (S.D. 0.0941). The $\alpha$ values for both conditions and the aggregated data from del.icio.us are situated in the interval $[1.732391 < \alpha < 2.249359]$. Figure 5 shows that both experimental conditions and the aggregated data from del.icio.us have similar exponents. Using the Monte Carlo sampling method recommended by Clauset et al. [5], for the 'No Tag Suggestion' condition $\alpha$ had a variance from 0.1266 to 0.1862 and in the 'tag suggestion' condition $\alpha$ had a variance from 0.1188 to 0.2097, thus leading the variation in the $\alpha$ of 'tag suggestion' and 'no tag suggestion' conditions to be statistically insignificant. For the 'no tag suggestion' condition, the variance of long-tail beginning for the power law fitting was 0.8348, while for the 'tag suggestion' condition it was 1.4805 [5]. Overall, our results show that the power law distribution, at least in the large, holds for both the 'tag suggestion' and 'no tag suggestion' condition, with no significant differences between their $\alpha$ parameters.

### 4.1.2 Kolmogorov-Smirnov Complexity

Determining whether a particular distribution is a 'good fit' for a power law is difficult, as most goodness-of-fit tests employ some sort of normal Gaussian assumption that is inappropriate for non-normal power law distributions. However, the Kolmogorov-Smirnov Test (abbreviated as the 'KS Test') can be employed for any distribution without implicit parametric assumptions and is thus ideal for use measuring goodness-of-fit of a given finite distribution to a power law function. Intuitively, given a reference distribution $P$ (perhaps produced by some well-known function like a power law) and a sample distribution $Q$ of size $n$, where one is testing the null hypothesis that $Q$ is drawn from $P$, then one simply compares the cumulative frequency of both $P$ and $Q$ and then the greatest discrepancy (the $D$-statistic) between the two distributions is tested against the critical value $D$-statistic for $n$, which varies per function. The null hypothesis is rejected if the $D$-statistic is greater than the critical value for $n$. For our experiment, we shall use $D < .1$ to indicate a significantly good fit with the power law distribution. While the technical details are beyond this paper (See work by Clauset et al. for details [5]).

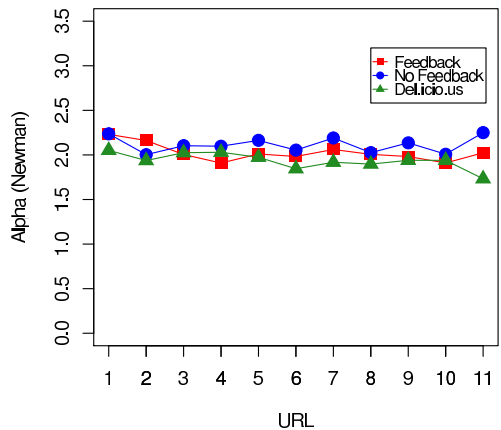

Figure 5: X axis depicts the URI used in the experiment, Y axis depicts the different $\alpha$ values

The KS test for all 11 tagged resources, testing both the 'tag suggestion' and 'no tag suggestion' condition is given in Figure 6. The average D statistic for the 'no tag suggestion' condition is 0.0313 (S.D. 0.0118), and for the "tag suggestion" condition the average $D$-statistic is 0.0724 (S.D. 0.0256). These results show that the power law function exhibited in both conditions is significant, the fit is closer for the 'no tag suggestion' condition than the "tag suggestion" condition. The $D$-statistic showed a range from 0.0170 to 0.0552 for "no tag suggestion' condition yet a range of 0.0428 to 0.1318 for 'tag suggestion,' implying that for some of the tag suggestion distributions a power-law is not even a remarkably good fit. Furthermore, since the $D$-statistic is based on the maximum discrepancy, this shows a larger discrepancy between the fitted power law for the 'tag suggestion' as opposed to the 'no tag suggestion' condition.

## 4.2 Influence of tag suggestion on the tag distribution

Given that the KS test shows that there is some difference, albeit subtle, between the power law distributions between the conditions, we need a more fine-grained way to tell if there is any difference in the distribution, particularly in the behavior of the long tail. A number of differing techniques, all previously widely used in tagging research, will be deployed to answer this question.

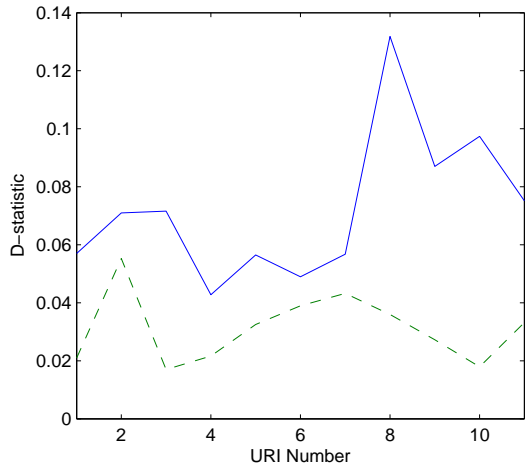

Figure 6: X axis depicts the URI used in the experiment, Y axis depicts the different D Statistics from the KS Test. The dotted line is the 'no tag suggestion' condition, while the solid line is the 'tag suggestion' condition.

### 4.2.1 Kullback Leibler Divergence

The Kullback-Leibler divergence (also known as *relative entropy*), which we abbreviate as 'KL divergence, ' can be used an intuitive information-theoretic measure of the distance between two distributions $P$ and $Q$. Unlike many other methods, it takes the entire distribution (in our case, the long tail is of particular interest) into account. Note that it is not a true metric as it is an asymmetric, however, it is a useful measure of the difference between two distributions as it is a non-negative, convex function, with well-known properties. The KL divergence is zero if and only if the two distributions are the same, otherwise a positive distance will result that will be larger the greater the divergence between the distributions. Intuitively in information theory, the KL divergence is the expected difference in bits required to encode to distribution $Q$ when using a code based on distribution $P$. The KL divergence between $P$ and $Q$ is given as:

$$D_{KL}(P||Q) = \sum_x P(x) log(\frac{P(x)}{Q(x)}) \qquad (3)$$

The KL divergence (using the "tag suggestion" condition for $P$ and the 'no tag suggestion' condition for $Q$) for each URI in the experiment are given in Figure 7. While some URIs (like number 6 and 7) have almost no difference between the "tag suggestion" and 'no tag suggestion' conditions, other URIs like number 11 have significant differences. This average KL divergence between the "tag suggestion" condition and 'no tag suggestion' condition is 0.1659 (S.D.0 0.0821 ). This is not insubstantial, and shows a considerable mismatch in the long tail. In particular, the long tail of the "tag suggestion" condition is often shorter than the "tag suggestion" condition, and the KL divergence takes this into account, while $\alpha$ does not.

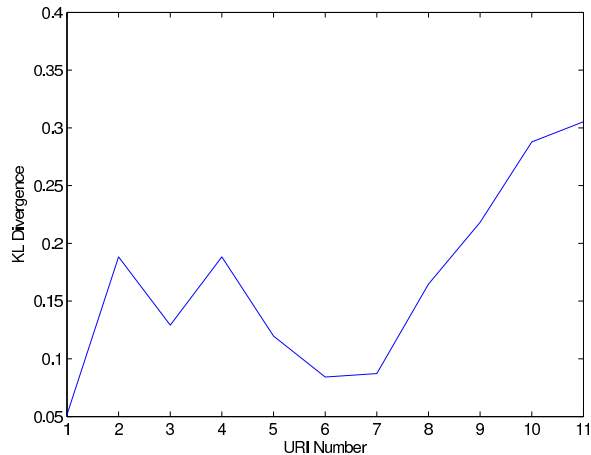

Figure 7: X axis depicts the URI used in the experiment, Y axis depicts the different KL Divergence values

### 4.2.2 Ranked frequency distribution

In order to observe the micro-behavior of the 'tag suggestion' and 'no tag suggestion' distributions, we investigate whether or not the tag suggestion tags are 'forced' higher in the distribution, so leading to the more a more sparse long tail and an exaggerated top of the distribution in the 'tag suggestion' condition. In order to provide a measurement of the number of suggested tags in the top of the distribution, the percentage of suggested tags that were found in the top 7 and top 10 tags were calculated. We compared the percentage of suggested tags in the top 7 and top 10 ranks for both conditions with del.icio.us. For this we assume that the 7 suggested tags provided by del.icio.us represent the top 7 tags in the ranked frequency distribution so that the percentage of suggested tags in the top 7 and top 10 ranks for del.icio.us is equal to 100%. We averaged the percentages for all URIs per experimental condition.

Figure 8 shows that for the percentage of suggested tags available in the top 7 rank for the 'tag suggestion' condition is 80.51% and for the 'no tag' suggestion condition 51.93%. This means that only half of the suggested tags can be found in the top 7 of the ranked frequency distribution in the 'no tag suggestion' condition. So in the 'tag suggestion' condition we observed more of the suggested tags in the top 10 rank of the ranked frequency distribution than in the 'no tag suggestion' condition. There is an influence of tag suggestions on the ranked position and the frequency of the suggested tags. Tag suggestions do influence the tag-resource distribution since on average half of the suggested tags do not at appear in the top 7 ranks, yet when suggested they do appear in the top 7 ranks. However, when users are not guided by tag suggestions and tag freely they still choose for themselves half of the tags that would have been otherwise suggested had they had a 'tag suggestion' mechanism available. Further we look at the availability of suggested tags in the top 10 as an indication how dispersed the suggested tags are in the ranked frequency distribution for both conditions. For the top 10 rank figure 8 shows that the percentage of suggested tags in the "tag suggestion" condition is 88.30% and for the "no tag suggestion" condition is 61.03%.

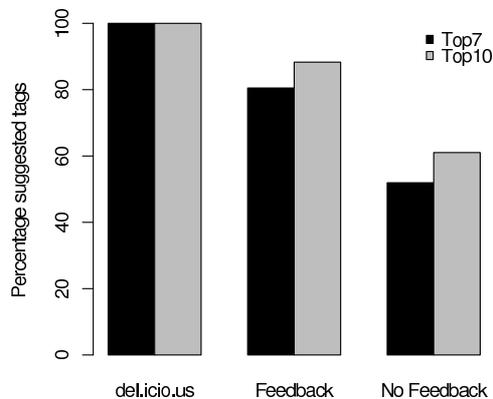

Figure 8: Ranked Frequency Distribution Repeating Suggested Tags

### 4.2.3 Matching and imitation rates

Another metric that measures the influence of tag suggestion on the tag distribution is the matching and imitation rate as proposed by Suchanek et al. [18]. The matching rate measure the proportion of applied tags that are available in the suggested tags. This metric provides insight in how the user is influenced by the tag suggestion provided by the tagging system. For our experiment the *matching rate* is being defined as :

$$mr(X) = \frac{\sum_{i=1}^{n} \mid T(X,i) \cap S(X) \mid}{\sum_{i=1}^{n} \mid T(X,i) \mid} \quad (4)$$

$X$ denotes the tag suggestion method that is being used in both our conditions. The 'tag suggestion' condition provides 7 suggested tags while the 'no tag suggestion' condition provided no suggested tags. For a given

URI, $T(X,i)$ denotes the set of tags at the $i$th tag entry and $S(X)$ denotes the suggested tags for that URI. For a tagging instance in which all tags are given by the suggested tags the matching rate will be 1.

The matching rate for the 11 URIs in the experiment and over the both conditions was calculated. The resulting matching rates can be found in Table 1. Condition 'no tag suggestion' serves as a reference point. The results in Table 1 show that users in the 'tag suggestion' condition are being influenced by the appearance of tag suggestions. The average matching rate for the 'tag suggestion' condition is 0.57 (S.D. 0.086) and for the no tag suggestion condition 0.35 (S.D. 0.068). The main drawback of the matching rate is that it can't account for the application of suggested tags when tag suggestion is absent.

Table 1: Matching rate

| URI No. | Tag Suggestion | No Tag Suggestion |
|---|---|---|
| 1 | 0.47 | 0.31 |
| 2 | 0.57 | 0.34 |
| 3 | 0.53 | 0.32 |
| 4 | 0.65 | 0.48 |
| 5 | 0.45 | 0.29 |
| 6 | 0.52 | 0.29 |
| 7 | 0.58 | 0.38 |
| 8 | 0.65 | 0.38 |
| 9 | 0.74 | 0.46 |
| 10 | 0.63 | 0.30 |
| 11 | 0.59 | 0.31 |

This ability to account for tag repetition even when the tag is missing is given by the imitation rate, defined as [18]:

$$\alpha_n(S) = \frac{prec_n(X,S) - prec_n(NONE,S)}{1 - prec_n(NONE,S)} \quad (5)$$

With :

$$prec_n = \frac{\sum_{i=1}^{n} \mid T(X,i) \cap S \mid [S(X,i) = S]}{\sum_{i=1}^{n} \mid T(X,i) \mid [S(X,i) = S]} \quad (6)$$

$prec_n$ defines the proportion of applied tags that are available in the single tag suggestion set $S$. Since the tags $S$ in our experiment is always static, $prec_n$ is equal to the calculation of the matching rate for the tag suggestion condition in Equation 4. $prec_n(NONE, S)$ defines the proportion of suggested tags that are available in the tags applied by the user when no tag suggestion is given. This is similar to the calculation of the matching rate for the 'no tag suggestion' condition. Therefore we can rewrite the imitation

rate as:
$$ir = \frac{mr(ConditionA) - mr(ConditionB)}{1 - mr(ConditionB)} \qquad (7)$$

Table 2 shows the imitation rates for the different experimental URIs. An imitation rate of 1 will denote full imitation. The results show that users tend to select suggested tags when the are available with a chance of 1 out of 3 with a mean imitation rate of 0.36 (S.D. 0.097).

Table 2: Imitation rate

| URI No. | Imitation Rate |
|---|---|
| 1 | 0.22 |
| 2 | 0.35 |
| 3 | 0.29 |
| 4 | 0.35 |
| 5 | 0.20 |
| 6 | 0.34 |
| 7 | 0.31 |
| 8 | 0.42 |
| 9 | 0.50 |
| 10 | 0.48 |
| 11 | 0.43 |

Combining this insight with our previous work, it appears that 'tag suggestion' condition causes more imitation in the top of the distribution and a 'shorter' long tail, in other words, a 'compression' of the basic power law distribution that both 'no tag suggestion' and 'tag suggestion' conditions generate. In the large, this can be observed by noticing that for the power law fitting, for the 'no tag suggestion' component, the variance of the long-tail cut-off for the power law fitting was 0.8348, while for the 'tag suggestion' condition it was 1.4805 [5]. In other words, the long tail of the 'tag suggestion' condition did not fit the power law function as well as the 'no tag suggestion' condition, and was cut off much earlier. It is because of this 'compression' caused by tag suggestions that 'tag suggestion' distributions do not fit power laws as well as 'no tag suggestion' distributions. Taking a 'scale-free' power law as an ideal stable tag distribution, rather counter-intuitively a simple tag suggestion scheme based on frequency may actually hurt rather than help the stabilization of tagging.

## 5 Conclusion

The research presented in this paper provides a first step that leads to a new interpretation of the accepted theories and models that explain the emergence of power laws in tagging systems. Common wisdom in tagging suggested that the power law was unlikely to form without tag suggestions. As put by Marlow, Boyd, and others, "A convergent folksonomy is likely to be generated when tagging is not blind," blind tagging being tagging without tag suggestions [10]. This does not appear to be the case from our experiment. The results show that a power law distribution emerges even when user are not guided by tag suggestions and tag freely without any tag suggestion from the tag-system. Moreover, the observed power law function fits *more closely* the behavior of users when the users are *not* given tag suggestions than when the users are given suggestions. This means that tag suggestions distorts the power law function that would already naturally occur when users tag without suggestion. The lack of effect of tag suggestions on the emergence of the power law distribution calls for a reinterpretation of current models of tagging. It appears that background knowledge is a much stronger influence than imitation.

Furthermore, these results are not entirely unexpected, but help clarify a number of experimental results from previous experiments in tagging. First, this result clarifies how the power law distribution was observed by Cattuto et al. even before del.icio.us began using tag suggestion via the tag interface [4]. Second, it also helps explain how the majority of users in Suchanek et al.'s experiment had a high matching rate, even when in their report-back most of them said they didn't use or even notice tag suggestions [18].

Our experiment does have a number of limitations, in particular our experiment should be extended to deal with more users as well as expert and non-expert users dealing with different kinds of subject matters they may have varying degrees of expertise in. In these kinds of situations, tag suggestions may have more of an influence on tagging behavior. Further research must be performed on determining the precise 'compression' behavior caused by tag suggestions and its effects on the long tail and top of the tag distributions. It appears, from our work here, that tag suggestions extend the top of the distribution (as shown in Fig. 4) while shortening the long tail.

Regardless, the cause of the emergence of the power law should be grounded in something else that a simple tag suggestion and imitation mechanism. Although the presented results indicate that some of the previous assumptions underlying the emergence of power laws do not hold, a power law distribution alone does not provide the necessary information needed to determine the role of tag suggestion on tag behavior. The results presented have merely confirmed the presence of a similar power law distribution for both 'tag suggestion' and 'no tag suggestion' conditions.

Another line of research that seems promising is to understand how human categorize in general, which could easily influence how they decide which tags to use to annotate web resources. For example, While the large amount of data on the web made it easy to develop simple mathematical models of human behavior, it seems that a crucial cognitive contributions of the user model are often ignored. What is missing is an understanding of the cognitive model and constraints of the tag selection process. Given the lack of attention seemingly paid suggested tags, it appears a cognitively-informed information retrieval approach may be an option for a new kind of tagging theory. An information-theoretic and cognitive understanding of the tag selection process is the next step understanding and extending our theories of tagging.

# 6 Acknowledgments

Dirk Bollen performed his research on the IBBT Project at CUO (Centre for User Experience Research) at the University of Leuven (KUL) and further elaborated on this study in the MyMedia project at the Eindhoven University of Technology. Harry Halpin received partial support from a Microsoft "Beyond Search" grant.